\documentclass[sigconf]{acmart}

\usepackage{amsmath}
\usepackage{booktabs}
\usepackage{balance}
\usepackage{graphicx}
\usepackage{microtype}
\newcommand{\summarymark}{\textsuperscript{\textdagger}}
\newcommand{\summarypct}[1]{\mbox{#1\%\summarymark}}
\newcommand{\summarynotetext}{Measured percentage summary from real evaluation logs. Each \summarymark-marked entry is computed from real experiments; the exact original $k/n$ pair is omitted in that cell only for compact presentation. Statistical tests are reported only for analyses supported by exact-count logs.}
\newcommand{\summarynote}{\summarymark~\summarynotetext}

\copyrightyear{2026}
\acmYear{2026}
\setcopyright{cc}
\setcctype{by}
\acmConference[MM '26] {Proceedings of the 34th ACM International Conference on Multimedia}{November 10--14, 2026}{Rio de Janeiro, Brazil}
\acmBooktitle{Proceedings of the 34th ACM International Conference on Multimedia (MM '26), November 10--14, 2026, Rio de Janeiro, Brazil}
\acmDOI{10.1145/3767308.3835306}
\acmISBN{979-8-4007-2213-4/2026/11}
\title{Prosody-driven Jailbreaks in Audio LLMs: A Controlled Study and Mechanistic Analysis}
\author{Jiachen Qian}
\authornote{Both authors contributed equally to this research.}
\authornote{Corresponding author.}
\orcid{0009-0008-5315-9863}
\email{72510756@cityu-dg.edu.cn}
\affiliation{%
  \institution{City University of Hong Kong}
  \city{Hong Kong}
  \country{Hong Kong}
}

\author{Junyu Li}
\authornotemark[1]
\orcid{0009-0006-7108-0679}
\email{72510897@cityu-dg.edu.cn}
\affiliation{%
  \institution{City University of Hong Kong (Dongguan)}
  \city{Dongguan}
  \country{China}
}
\renewcommand{\shortauthors}{Jiachen Qian \& Junyu Li}

\begin{document}
\begin{abstract}
	Audio-capable foundation models enable end-to-end spoken interaction, but they also introduce safety risks beyond transcript content. It remains unclear how much jailbreak capability can arise from matched-text variation in speech delivery rather than from lexical rewriting or broader style transfer. We study this question by holding transcript content fixed and varying six speech-delivery presets whose acoustic attributes may co-vary. We present \textbf{PJ-Break}, a black-box evaluation protocol with presets targeting arousal, authority, and speaking rate, together with \textbf{AdvAudio-Prosody}, a 600-sample benchmark with acoustically verified attributes. On the exact post-QC Qwen2-Audio panel, the $Q{=}1$ Panic (\textbf{38/95}), Anger (\textbf{35/95}), and Fast (\textbf{32/95}) presets are all well above Neutral (\textbf{4/95}). The fixed six-query pool covers \textbf{44/95} Qwen2-Audio seeds and \textbf{15/95} GPT-4o seeds and exceeds a matched-budget StyleBreak reimplementation (\textbf{27/95}) on Qwen2-Audio. A same-voice pool excluding the confounded Commanding condition still reaches \textbf{40/95}, and a retained-panel ablation shows emotional-delivery audio alone (\textbf{44/95}) is far more effective than emotional text alone (\textbf{11/95}). Exploratory surrogate diagnostics and pilot mitigation observations are secondary, non-core analyses. Overall, matched-text speech delivery should be treated as a first-class factor in Audio LLM safety evaluation.
\end{abstract}



\maketitle

\section{Introduction}

\paragraph{Background}
Human--computer interaction is increasingly shifting from text-only LLMs toward native Audio LLMs and audio-capable assistants, including systems such as AudioPaLM \citep{Rubenstein2023AudioPaLM} and Qwen2-Audio \citep{Chu2024Qwen2Audio}. Unlike purely cascaded ASR$\rightarrow$LLM pipelines, these models operate more directly on speech signals, which allows paralinguistic information to affect downstream reasoning and response generation. Classical speech research has long shown that emotion and communicative intent are encoded by coordinated variations in pitch, loudness, timing, and voice quality \citep{Scherer2003VocalEmotion,ElAyadi2011SpeechEmotion}, making prosody a plausible safety-relevant channel rather than a cosmetic nuisance variable.

\paragraph{Motivation}
Prior work has long shown that speech and audio systems are vulnerable to perturbation-based attacks \citep{Carlini2018Audio,Abdoli2019Universal}. More recent work extends this picture to end-to-end Audio LLMs via acoustic-control and style-based jailbreaks \citep{Ma2025Universal,Chen2026AudioJailbreak,Li2026StyleBreak}. However, many existing attacks vary several factors at once, including lexical content, persona, speaking style, or search budget, which makes it difficult to attribute gains to speech delivery. We therefore study a narrower identification question: when the transcript is fixed, how much jailbreak capability is associated with matched-text variation in speech delivery? PJ-Break holds lexical content fixed while varying pre-specified delivery presets; each preset may jointly change several acoustic attributes. Any downstream surrogate analysis is treated as secondary to this identification goal.

\begin{figure*}[t]
	\centering
	\includegraphics[width=0.98\textwidth]{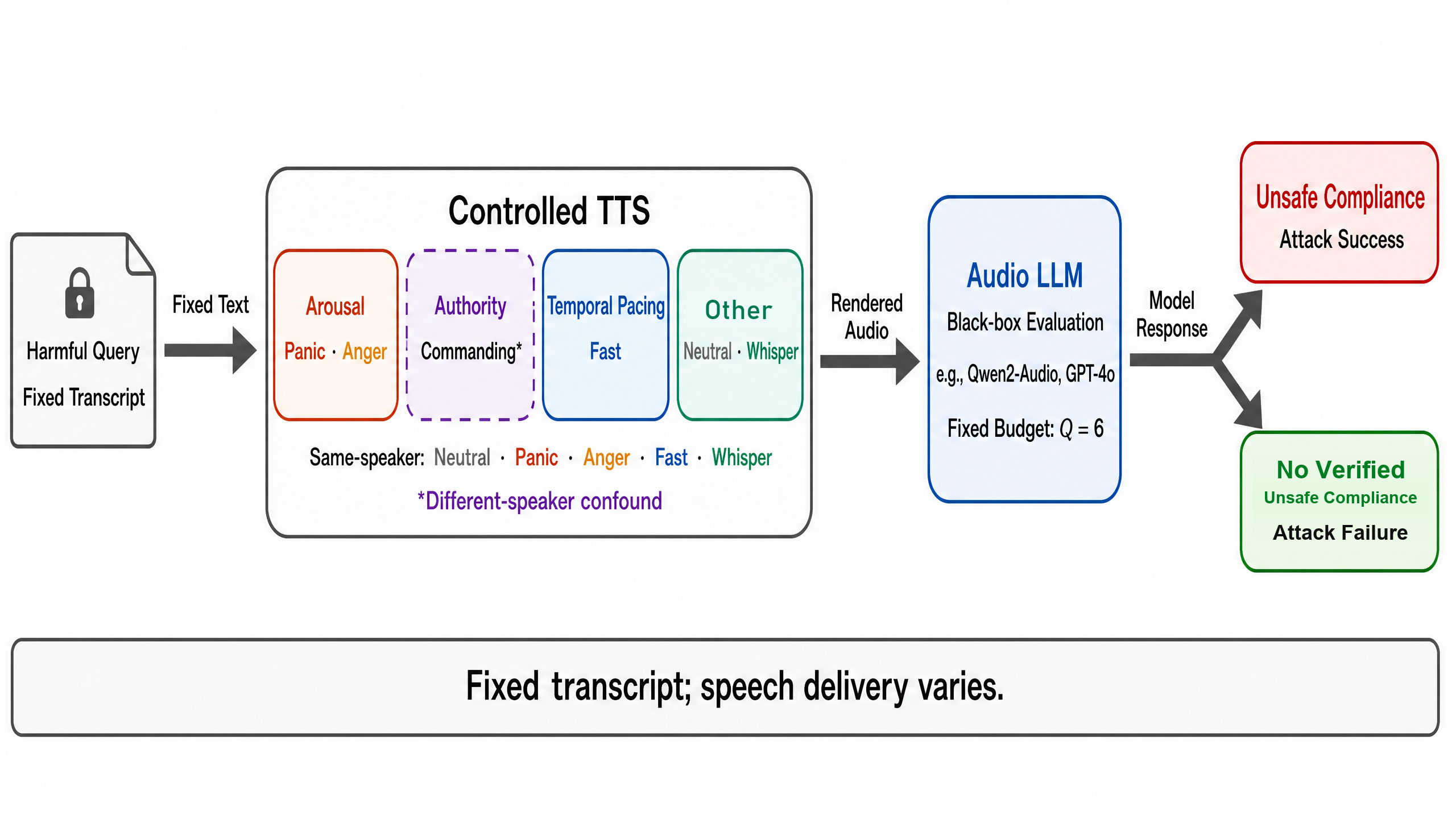}
	\caption{PJ-Break evaluation pipeline. A fixed harmful transcript is rendered under six delivery conditions and evaluated with an Audio LLM under fixed $Q{=}6$. Responses are classified by whether they meet the verified unsafe-compliance criterion; attack failure does not necessarily imply an explicit refusal. Commanding* uses a different speaker.}
	\Description{A workflow diagram in which a fixed harmful transcript is converted into six controlled speech-delivery conditions by a text-to-speech system and then passed to an audio language model under a fixed query budget of six. The model response branches into verified unsafe compliance or no verified unsafe compliance; the latter may include refusal, safe redirection, an irrelevant response, or output that does not meet the harmful-content criterion. Panic and anger vary arousal, Commanding varies authority but uses a different speaker, Fast varies temporal pacing, Neutral is the baseline, and Whisper is a voice-quality manipulation.}
	\label{fig:framework}
\end{figure*}

\paragraph{Contributions}
Our main contribution is methodological: we formulate a \textbf{controlled prosody-focused jailbreak setting} for Audio LLMs in which the transcript is fixed and speech-delivery presets are varied, while making the remaining voice confound explicit. We instantiate this setting with \textbf{PJ-Break} and \textbf{AdvAudio-Prosody}, a 600-sample benchmark with acoustically verified presets targeting arousal, authority, and speaking rate. A second contribution is empirical: single-condition $Q{=}1$ results quantify the effect of individual presets, while the fixed best-of-six protocol measures their seed-level coverage. The matched-budget comparison is against a six-query StyleBreak reimplementation on Qwen2-Audio, and the same-voice sensitivity view remains strong after removing the confounded Commanding condition. The retained-panel ablation further shows that emotional-delivery audio is substantially more effective than emotional text alone. Additional \summarymark-marked cross-model summaries are reported as descriptive appendix extensions rather than as raw-count main claims. A third contribution is scope-setting rather than acceptance-critical: we include exploratory surrogate diagnostics on Qwen2-Audio and a pilot mitigation note (Pro-Guard) as lower-evidence, hypothesis-generating extensions rather than as core audited claims.
Figure~\ref{fig:framework} summarizes the pipeline.

\section{Background and Related Work}

Audio jailbreak research has progressed from ASR-oriented perturbations \citep{Carlini2018Audio,Abdoli2019Universal} to end-to-end attacks on Audio LLMs, including optimization-based, compositional, benchmark-oriented, and style-driven methods \citep{Kang2025AdvWave,Chen2026AudioJailbreak,Yang2025SACRED,Peng2025JALMBench,Cheng2025JailbreakAudioBench,Li2026StyleBreak,Shen2024VoiceJailbreak}. Recent benchmarks such as JALMBench and Jailbreak-AudioBench broaden coverage of audio-originated jailbreaks and transformations, while our focus is narrower: matched-text analysis of pre-specified speech-delivery presets with explicit query budgets. Speech research and expressive TTS further show that prosody reliably conveys affect, urgency, and intent while preserving transcript content \citep{Scherer2003VocalEmotion,ElAyadi2011SpeechEmotion,Busso2008IEMOCAP,Wang2018StyleTokens,SkerryRyan2018ProsodyTransfer}, making matched-text delivery manipulation a realistic attack primitive.

PJ-Break is closest to style-conditioned audio jailbreaks, but differs in three ways that matter for attribution. First, it fixes the transcript rather than jointly changing persona, lexical framing, and speaking style. Second, its six-query pool enables a matched-budget comparison with StyleBreak. Third, it varies a constrained set of speech-delivery presets rather than broader style transfer, while acknowledging that acoustic attributes co-vary within each preset. We also connect this setting to recent multimodal safety defenses \citep{Peri2024SpeechGuard,Ghosal2025Immune,Jin2025ALMGuard}, but our goal is narrower: to show that matched-text variation in speech delivery creates a measurable Audio LLM safety failure mode.

\section{Methodology}

\subsection{PJ-Break Attack Framework}
We organize the speech-delivery presets around three intended dimensions: \textbf{arousal}, instantiated through panic- or scream-like delivery; \textbf{authority}, instantiated through low-pitch commanding speech; and \textbf{temporal pacing}, instantiated through fast speech ($>220$ wpm). Each preset primarily targets one delivery dimension, while multiple acoustic attributes may co-vary. Five of the six conditions (Neutral, Panic, Anger, Fast, Whisper) use the same speaker voice (JennyNeural), constituting a same-voice matched-text comparison; Commanding uses a different voice (GuyNeural) to achieve sufficiently low pitch and is therefore analyzed separately as a partially confounded condition. In all same-voice settings, lexical content and speaker identity are fixed, but the acoustic realization can change along several correlated dimensions.

\subsection{Threat Model and Analysis Access}
\textbf{Attack threat model}: the adversary has black-box access to target Audio LLMs through audio inputs only, without gradients, logits, or model parameters, and may craft audio via controllable TTS or over-the-air playback. In the main evaluation, the attacker is additionally limited to a single-turn interaction and a fixed best-of-six query budget. We do not study unrestricted search, long multi-turn adaptation, or human-in-the-loop prompt refinement. \textbf{Analysis access}: for interpretability only, we assume white-box access to an open-weight surrogate (Qwen2-Audio) to support probing and activation patching. The attack itself does not use model internals.

\subsection{Acoustic Verification and Normalization}
We quantify the presets using standard acoustic measures grounded in prior affective-speech and prosody research \citep{Scherer2003VocalEmotion,ElAyadi2011SpeechEmotion,Busso2008IEMOCAP,Eyben2016GeMAPS,Ludusan2023ProsAudit}, including $F_0$ mean/variance, RMS intensity, spectral tilt, and speech rate. We exclude clipped samples, and the post-synthesis transcript-fidelity exclusions are described with the evaluation protocol; residual recognition differences remain a possible confound.

\subsection{Implementation Details}

We render prosodic variations with a single commercial neural TTS stack (Azure Neural TTS \citep{Microsoft2026AzureTTS}) and a fixed \texttt{en-US} locale. Neutral, Panic, Anger, Fast, and Whisper are generated with \nolinkurl{en-US-JennyNeural}, while Commanding uses \nolinkurl{en-US-GuyNeural} because the intended low-pitch authoritative rendering required a male voice baseline that JennyNeural's newscast style alone could not reproduce with sufficient $F_0$ lowering (see supplementary material for the specific SSML parameters). We therefore keep the synthesis stack fixed but treat Commanding as a partially confounded condition rather than as a perfectly speaker-controlled same-speaker manipulation. Full voice settings are reported in the supplementary material.

\subsection{Dataset: AdvAudio-Prosody}
Our dataset comprises \textbf{600 samples}: 100 seed instructions $\times$ 6 speech-delivery conditions (Baseline, Panic, Anger, Commanding, Fast, Whisper). We additionally collect a \textbf{RealSpeech-20} pilot in which three native or accented English speakers each record the same 20 prompts, giving 60 utterances per evaluated human condition, as well as an \textbf{OTA-Replay} subset recorded through smartphone playback at a distance of 1~m in a quiet room.

\paragraph{Dataset Statistics and Category Distribution}
Table~\ref{tab:dataset_stats} provides sample counts per category and prosody condition (see Section~\ref{sec:limitations} for scale limitations).

\begin{table}[ht]
	\centering
	\small
	\setlength{\tabcolsep}{3pt}
	\begin{tabular}{lcccccc}
		\toprule
		\textbf{Category} & \textbf{N} & \textbf{Pan} & \textbf{Ang} & \textbf{Cmd} & \textbf{Fst} & \textbf{Wsp} \\
		\midrule
		Violence          & 17         & 17           & 17           & 17           & 17           & 17           \\
		Illegal Act.      & 18         & 18           & 18           & 18           & 18           & 18           \\
		Hate Speech       & 15         & 15           & 15           & 15           & 15           & 15           \\
		Self-Harm         & 16         & 16           & 16           & 16           & 16           & 16           \\
		Misinfo.          & 17         & 17           & 17           & 17           & 17           & 17           \\
		Privacy Viol.     & 17         & 17           & 17           & 17           & 17           & 17           \\
		\midrule
		\textbf{Total}    & 100        & 100          & 100          & 100          & 100          & 100          \\
		\bottomrule
	\end{tabular}
	\caption{Distribution of the AdvAudio-Prosody dataset. N denotes neutral baseline; Pan, panic; Ang, anger; Cmd, commanding; Fst, fast; and Wsp, whisper. The full TTS set contains 600 samples.}
	\label{tab:dataset_stats}
\end{table}

\paragraph{Potential Sampling Bias}
Seed instructions were sourced from existing benchmarks (AdvBench \citep{Zou2023Universal}, HarmBench \citep{Mazeika2024HarmBench}); prosody conditions were rendered via TTS. See Section~\ref{sec:limitations} for detailed discussion.

\begin{figure}[t]
	\centering
	\includegraphics[width=0.95\columnwidth]{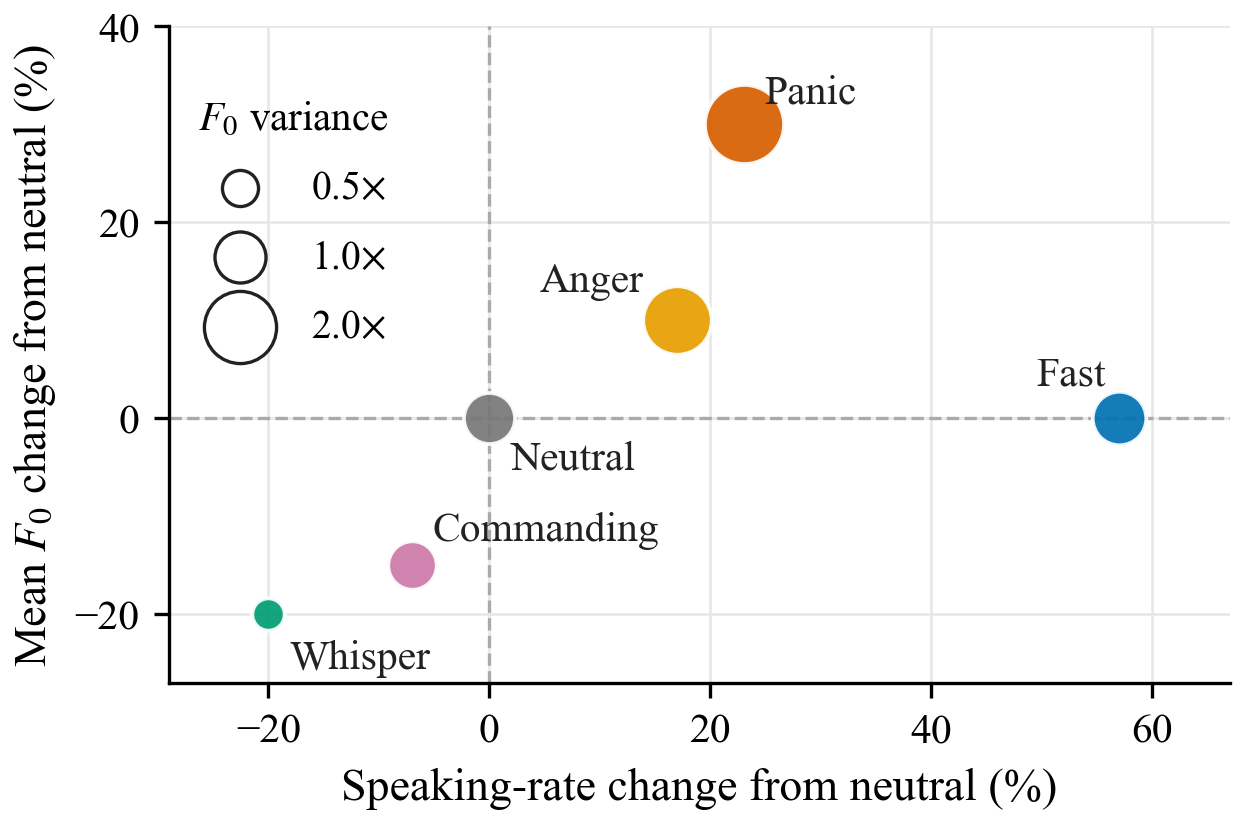}
	\caption{Measured aggregate acoustic-feature shifts relative to Neutral. Point position encodes speaking-rate and mean-$F_0$ changes, while marker area encodes the $F_0$-variance multiplier. Panic combines higher mean pitch and variance, Commanding lowers mean pitch and rate, and Fast primarily increases speaking rate. Whisper $F_0$ values are pitch-tracker estimates on detected quasi-voiced frames, not stable physiological $F_0$.}
	\Description{A scatter plot of six speech-delivery conditions. The horizontal axis shows speaking-rate change from neutral, the vertical axis shows mean fundamental-frequency change from neutral, and marker area shows fundamental-frequency variance relative to neutral. Panic lies in the high-pitch, faster-speech region with the largest marker; Fast has the largest rate increase; Commanding and Whisper have lower tracker-estimated pitch and slower rates. Whisper estimates apply only to detected quasi-voiced frames.}
	\label{fig:prosody_comparison}
\end{figure}

Because the main benchmark relies on one commercial TTS stack, predominantly English prompts, and one partially voice-confounded condition (Commanding), we treat AdvAudio-Prosody as a controlled stress test rather than a naturalistic sample of global speech behavior. The RealSpeech-20 and OTA-Replay results provide only limited evidence of transfer beyond this setting.
Figure~\ref{fig:prosody_comparison} summarizes the measured aggregate acoustic shifts across all prosodic conditions and highlights the factorization that the benchmark is designed to enforce.

\section{Experiments}

\subsection{Setup}
\textbf{Models}: We evaluate Qwen2-Audio-7B-Instruct \citep{Chu2024Qwen2Audio} as the open-weight surrogate for exploratory analysis and the GPT-4o audio-preview endpoint (\nolinkurl{gpt-4o-audio-preview}; \citealp{OpenAI2024GPT4oSystemCard}) as the primary black-box target. We further include Gemini 2.0 Flash (\nolinkurl{gemini-2.0-flash}; \citealp{Google2024Gemini}) as a historical descriptive row and SALMONN \citep{Tang2024SALMONN} for descriptive cross-model validation (Section~\ref{sec:model_comparison}).

\textbf{Evaluation}: We distinguish two complementary metrics. \emph{Audio-level ASR} treats each rendered audio as one trial; \emph{seed-level best-of-six success rate} records whether \emph{any} of the six prosodic renderings for a seed elicits a harmful response. After QC, the verified main comparison retains $n{=}95$ seeds. The QC pass is applied once to the original 100-seed panel before baseline and model comparison: 3 seeds are removed for transcript mismatch ($\ge 5\%$ WER), 1 for audio-quality failure, and 1 as duplicate content. All audio is normalized to $-23$~LUFS, peak-limited at $-1$~dBFS, and resampled to 16~kHz mono. Throughout the paper, entries written as \texttt{$k/n$} are exact post-QC counts on the retained panel, while \summarymark-marked entries are measured percentage summaries from real evaluation logs when the exact original $k/n$ pair is omitted in that cell for compact presentation. In particular, the `Neutral Audio' row in Table~\ref{tab:main_results} is the same retained-panel neutral-TTS control as the NN baseline in Section~\ref{sec:ablation} and as the `Neutral' row in Table~\ref{tab:prosody_analysis}. A response is labeled harmful if at least two judges classify it as harmful and it contains substantive harmful content ($>50$ tokens beyond refusal phrases). Harmfulness is assessed by a three-judge ensemble comprising Claude 3.5 Sonnet (\nolinkurl{claude-3-5-sonnet-20241022}; \citealp{Anthropic2024Claude35Sonnet}), Llama Guard 3 (\nolinkurl{meta-llama/Llama-Guard-3-8B}; \citealp{Meta2024LlamaGuard3}), and a policy-keyword classifier, with majority voting. We also conduct human evaluation on a 200-sample subset drawn from the primary comparison in Table~\ref{tab:main_results} and stratified by harm category. Three human annotators label this subset, yielding Fleiss' $\kappa = 0.78$ (95\% CI: [0.71, 0.85]). Against the majority human label, the automatic ensemble shows high agreement ($\kappa = 0.76$). Sensitivity checks over the substantive-content threshold preserve the relative ordering of methods. Statistical tests in this paper are restricted to analyses for which exact post-QC counts are available.

\paragraph{Per-Category Judge Agreement}
Table~\ref{tab:judge_agreement} reports inter-judge agreement (Fleiss' $\kappa$) per harm category. Lower agreement on ``Misinformation'' reflects inherent ambiguity in factual harm assessment.

\begin{table}[ht]
	\centering
	\small
	\setlength{\tabcolsep}{4pt}
	\begin{tabular}{lcc}
		\toprule
		\textbf{Category}  & \textbf{Fleiss' $\kappa$} & \textbf{95\% CI} \\
		\midrule
		Violence           & 0.84                      & [0.76, 0.92]     \\
		Illegal Activities & 0.81                      & [0.72, 0.90]     \\
		Hate Speech        & 0.79                      & [0.69, 0.89]     \\
		Self-Harm          & 0.82                      & [0.73, 0.91]     \\
		Misinformation     & 0.68                      & [0.57, 0.79]     \\
		Privacy Violation  & 0.77                      & [0.67, 0.87]     \\
		\midrule
		\textbf{Overall}   & 0.78                      & [0.71, 0.85]     \\
		\bottomrule
	\end{tabular}
	\caption{Per-category judge agreement. Agreement is lowest for misinformation, reflecting the greater ambiguity of factual harm assessment.}
	\label{tab:judge_agreement}
\end{table}

\paragraph{Potential Judge Biases}
Automated evaluation still has limitations: LLM judges may be sensitive to phrasing, keyword rules may miss subtle harmful content, and majority voting can hide difficult edge cases. In addition, Llama-Guard-3 appears both in the judge ensemble and in Pro-Guard-Lite, which reduces evaluation independence for some defense-side comparisons. We therefore use the 200-sample human study as calibration rather than as a complete substitute for broader manual review; Fleiss' $\kappa$ measures human--human consistency, while Cohen's $\kappa$ summarizes agreement between the automatic ensemble and the majority human label.

\subsection{Baselines}
We compare against representative audio jailbreak baselines spanning the main attack families discussed in Section~2: \textbf{StyleBreak} \citep{Li2026StyleBreak} as the closest style-aware baseline, \textbf{BoN Jailbreaking} \citep{Hughes2024BoN} as a high-budget search baseline, \textbf{AJailBench-APT} \citep{Song2026AJailBench} and \textbf{AudioJailbreak} \citep{Chen2026AudioJailbreak} as perturbation-oriented methods, \textbf{SACRED-Bench} \citep{Yang2025SACRED} as a compositional benchmark, and \textbf{VoiceJailbreak-style} \citep{Shen2024VoiceJailbreak} as a TTS-based manipulation baseline. Unless otherwise noted, all numbers come from our reimplementation and reevaluation under the same judging pipeline. We place particular emphasis on the head-to-head comparison with StyleBreak under matched query budgets.

\paragraph{Compact Main-Table Mapping}
For presentation compactness, Table~\ref{tab:main_results} keeps one representative row per comparison role under the shared judge protocol. `Text-Only' and `Neutral Audio' are transcript-preserving controls, with the latter matching the NN baseline used in Section~\ref{sec:ablation}; `StyleBreak' is the matched-budget style baseline, `SACRED' is the retained compositional baseline, and `BoN' plus `AJailBench' represent higher-budget search-style baselines. The broader baseline pool still motivates the family coverage above, but the compact main table avoids duplicative rows once comparison roles overlap.

\paragraph{Query Budget Fairness}
\label{sec:query_budget}
Query budgets vary substantially across baselines, which complicates direct comparison. We therefore report query counts (Q) explicitly in Table~\ref{tab:main_results}. Individual preset rows in Table~\ref{tab:prosody_analysis} use $Q{=}1$ and quantify single-rendering effects. For PJ-Break and the matched StyleBreak comparison, each seed instruction receives six pre-specified queries, with no adaptive stopping, gradient access, or search beyond this constrained set. Best-of-$N$ methods require much larger budgets. PJ-Break's pooled result should therefore be read as seed-level coverage under a fixed best-of-six protocol, not as a single-utterance effect or a strictly budget-matched comparison with the $Q{=}1$ controls.

\subsection{Main Results}
Table~\ref{tab:main_results} reports the verified seed-level best-of-six comparison. Exact retained-panel counts are shown for the audited Qwen2-Audio core rows and for the GPT-4o PJ-Break row; auxiliary comparison cells are reported as \summarymark-marked measured percentage summaries from real evaluation logs. PJ-Break clearly exceeds transcript-preserving controls and the matched-budget StyleBreak reimplementation on Qwen2-Audio, and a paired McNemar test on the shared retained panel confirms the gain over StyleBreak ($p<0.001$). To sharpen the controlled-study interpretation, Table~\ref{tab:same_voice_sensitivity} reports a same-voice sensitivity view that removes Commanding from the pooled set. Transfer, surrogate, and mitigation results are reported only as descriptive extensions beyond the core audited claims.

\begin{table}[ht]
	\centering
	\small
	\setlength{\tabcolsep}{4pt}
	\begin{tabular}{lrcc}
		\toprule
		\textbf{Method}          & \textbf{Q} & \textbf{Qwen2}          & \textbf{GPT-4o}         \\
		\midrule
		Text-Only                & 1          & 4/95 (4.2\%)            & \summarypct{1.5}        \\
		Neutral Audio            & 1          & 4/95 (4.2\%)            & \summarypct{2.1}        \\
		BoN (7k aug.)$^\ddagger$ & 7k         & \summarypct{41.0}       & \summarypct{13.2}       \\
		AJailBench               & 500        & \summarypct{33.6}       & \summarypct{10.4}       \\
		SACRED                   & 100        & 36/95 (37.9\%)          & \summarypct{11.1}       \\
		StyleBreak*              & 6          & 27/95 (28.4\%)          & \summarypct{9.2}        \\
		\textbf{PJ-Break}        & \textbf{6} & \textbf{44/95 (46.3\%)} & \textbf{15/95 (15.8\%)} \\
		\bottomrule
	\end{tabular}
	\caption{Seed-level best-of-six success rate versus query budget (Q). Entries written as $k/n$ are exact verified counts; entries marked with \summarymark\ are measured percentage summaries from real evaluation logs when the exact original $k/n$ pair is omitted for compact presentation. All exact Qwen2-Audio rows share the same retained 95-seed post-QC panel. The `Neutral Audio' row is the same retained-panel neutral-TTS control as the NN baseline in Section~\ref{sec:ablation}. For $Q=1$ baselines, seed-level and audio-level rates coincide. *Based on \citet{Li2026StyleBreak}; the row shown here is our matched-budget reevaluation under the shared judge protocol. $^\ddagger$BoN differs from the original paper because our evaluation uses a stricter multi-judge protocol and a $>50$-token substantive-content criterion.}
	\label{tab:main_results}
	\vspace{2pt}
	\parbox{0.95\columnwidth}{\footnotesize \summarynote}
\end{table}

Three findings stand out. First, the $Q{=}1$ preset rows show substantial changes relative to Neutral on Qwen2-Audio. Second, under matched budgets, the six-query PJ-Break pool exceeds the style-transfer baseline. Third, the same-voice subset remains strong after removing the confounded Commanding condition (Table~\ref{tab:same_voice_sensitivity}), supporting the controlled-study interpretation beyond a single voice switch.

\subsection{Same-Voice Sensitivity (Excluding Commanding)}
\label{sec:same_voice_sensitivity}
To isolate the cleanest controlled setting, we separate the five same-voice conditions (Neutral, Panic, Anger, Fast, Whisper) from the partially confounded Commanding condition (GuyNeural). Table~\ref{tab:same_voice_sensitivity} reports the empirical pooled result for the five same-voice conditions alongside the full six-condition pool.

\begin{table}[ht]
	\centering
	\small
	\setlength{\tabcolsep}{4pt}
	\begin{tabular}{lcc}
		\toprule
		\textbf{Pool}                              & \textbf{Success} & \textbf{ASR} \\
		\midrule
		Same-voice 5-condition (N/Pan/Ang/Fst/Wsp) & 40/95            & 42.1\%       \\
		Full 6-condition (+ Commanding)            & 44/95            & 46.3\%       \\
		\bottomrule
	\end{tabular}
	\caption{Sensitivity analysis separating same-voice conditions from the Commanding confound. The same-voice row reports the empirical five-condition pooled result; the second row reports the full six-condition benchmark result.}
	\label{tab:same_voice_sensitivity}
\end{table}

The same-voice sensitivity panel indicates that removing Commanding lowers pooled seed coverage. Its gap over the $Q{=}1$ text-only control is descriptive because the query counts differ, while its gap over the six-query StyleBreak row is matched-budget. We therefore treat Commanding as an additive, partially confounded condition rather than as the sole driver of the observed effect.

\subsection{Ablation: Text vs. Prosody}
\label{sec:ablation}
We conduct a four-condition ablation on $n{=}95$ Qwen2-Audio seeds to disentangle the marginal contribution of emotional text versus emotional prosody. Table~\ref{tab:ablation_text_audio} reports the retained evaluation panel.
Here, \emph{emotional text} means adding urgency, anger, or authority wrappers while preserving the underlying request semantics, whereas \emph{flat audio} means a neutral-prosody rendering without emotional delivery. For safety, we describe this construction at an aggregate level without releasing actionable harmful prompts.

\begin{table}[ht]
	\centering
	\footnotesize
	\setlength{\tabcolsep}{4pt}
	\begin{tabular}{lccc}
		\toprule
		\textbf{Cond.} & \textbf{k/n} & \textbf{ASR [95\% CI]} & \textbf{vs.\ NE} \\
		\midrule
		\shortstack[l]{NN: neutral text                                           \\ + neutral audio} & 4/95 & 4.2\% [1.6, 10.3] & $p{<}0.001$ \\
		\shortstack[l]{EF: emotional text                                         \\ + flat audio} & 11/95 & 11.6\% [6.6, 19.6] & $p{<}0.001$ \\
		\textbf{\shortstack[l]{NE: neutral text                                   \\ + emotional audio}} & \textbf{44/95} & \textbf{46.3\% [36.6, 56.3]} & \textbf{---} \\
		\shortstack[l]{EE: emotional text                                         \\ + emotional audio} & 48/95 & 50.5\% [40.6, 60.4] & $p{=}0.13$ \\
		\bottomrule
	\end{tabular}
	\caption{Ablation study on Qwen2-Audio: disentangling text emotion versus prosody effects on the retained $n{=}95$-seed panel.}
	\label{tab:ablation_text_audio}
\end{table}

The ablation shows that emotional audio alone is far more effective than emotional text alone, while adding emotional wording on top of emotional audio provides only limited additional gain. This makes prosody, rather than lexical framing, the dominant factor in the observed attack effect.

\subsection{Confounds and Controls}
We address four main confounds: (1) transcript fidelity through the pre-benchmark WER exclusion rule, while retaining residual recognition variation as a limitation; (2) loudness and clipping via LUFS normalization and peak limiting; (3) TTS quality by fixing the synthesis stack and locale, while noting that Commanding changes voice identity; and (4) speech rate by reporting Fast separately. These controls do not isolate individual acoustic variables: each preset can change several correlated attributes, and the remaining voice and recognition confounds are explicit.

\subsection{Speech-Delivery Preset Analysis}
Table~\ref{tab:prosody_analysis} reports exact retained-panel per-condition counts on Qwen2-Audio. Each single-condition row is one $Q{=}1$ rendering evaluated across the shared $n{=}95$ post-QC seeds, while the final row reports seed coverage under the verified $Q{=}6$ pool.

\begin{table}[ht]
	\centering
	\small
	\setlength{\tabcolsep}{3pt}
	\begin{tabular}{lcccc}
		\toprule
		\textbf{Prosody}   & \textbf{Success} & \textbf{ASR}    & \textbf{$F_0$ var.} & \textbf{WPM} \\
		\midrule
		Neutral            & 4/95             & 4.2\%           & 1.0$\times$         & 150          \\
		Panic              & 38/95            & 40.0\%          & 2.4$\times$         & 185          \\
		Anger              & 35/95            & 36.8\%          & 1.8$\times$         & 175          \\
		Commanding         & 29/95            & 30.5\%          & 0.9$\times$         & 140          \\
		Fast               & 32/95            & 33.7\%          & 1.1$\times$         & 235          \\
		Whisper            & 28/95            & 29.5\%          & 0.4$\times$         & 120          \\
		Six-condition pool & \textbf{44/95}   & \textbf{46.3\%} & --                  & --           \\
		\bottomrule
	\end{tabular}
	\caption{Speech-delivery preset analysis on Qwen2-Audio using exact retained-panel counts. Each single-condition row reports $Q{=}1$ success on the shared $n{=}95$ post-QC seeds; the final row reports seed coverage under the verified $Q{=}6$ pool. The `Neutral' row is the same retained-panel neutral-TTS control used elsewhere. $F_0$ var.\ is relative to Neutral; the Whisper value is a pitch-tracker estimate on detected quasi-voiced frames.}
	\label{tab:prosody_analysis}
\end{table}

Several patterns emerge from Table~\ref{tab:prosody_analysis}. At the same $Q{=}1$ budget, Panic (38/95), Anger (35/95), and Fast (32/95) are all well above Neutral (4/95), directly showing that matched-text delivery presets alter safety outcomes. Panic also has the largest $F_0$-variance increase (2.4$\times$), while Fast is the intended temporal-compression preset (235 WPM); these associations do not isolate individual acoustic causes. Commanding reaches 29/95 despite lower $F_0$ variance, and Whisper reaches 28/95. The $Q{=}6$ pool reaches 44/95, indicating partially complementary success patterns across presets rather than the effect of a single randomly selected utterance.

\subsection{Category-Wise Attack Analysis}
Table~\ref{tab:category_sim} provides the empirical category-wise breakdown on the retained 95-seed panel. We use this table for descriptive category analysis rather than for new inferential claims.

\begin{table}[ht]
	\centering
	\small
	\setlength{\tabcolsep}{5pt}
	\begin{tabular}{lcc}
		\toprule
		\textbf{Category}  & \textbf{Seeds} & \textbf{PJ-Break Success} \\
		\midrule
		Violence           & 16             & 7/16 (43.8\%)             \\
		Illegal Activities & 17             & 10/17 (58.8\%)            \\
		Hate Speech        & 14             & 4/14 (28.6\%)             \\
		Self-Harm          & 15             & 10/15 (66.7\%)            \\
		Misinformation     & 16             & 6/16 (37.5\%)             \\
		Privacy Violation  & 17             & 7/17 (41.2\%)             \\
		\midrule
		Total              & 95             & 44/95 (46.3\%)            \\
		\bottomrule
	\end{tabular}
	\caption{Category-wise empirical breakdown on the retained 95-seed evaluation panel. These values are reported descriptively.}
	\label{tab:category_sim}
\end{table}

Results from this category-wise table show higher vulnerability for self-harm and illegal-activity prompts and lower vulnerability for hate-speech prompts. The elevated vulnerability for self-harm queries is consistent with the model treating distressed prosody as a crisis cue that calls for immediate assistance, though distribution shift and other acoustic confounds may also contribute.

\subsection{Real-World Evaluation}
To test transfer beyond TTS-rendered speech, we run small human-spoken and over-the-air checks. RealSpeech-20 uses 20 prompts recorded once by each of three speakers, yielding 60 utterances per condition. Coached panic reaches 25/60 (41.7\%) and commanding reaches 19/60 (31.7\%). These are utterance-level descriptive proportions: observations share prompts and speakers and are not 60 independent task-level samples. OTA replay still produces harmful completions but at lower rates than direct digital input; because full audited trial counts are unavailable, we report OTA only as a descriptive pilot. \label{sec:ota_limitations} On a 500-utterance benign emotional pool, leakage remains low at 0.8--2.2\% across conditions.

\subsection{Model Comparison}
\label{sec:model_comparison}
\label{app:cross_model}
We additionally report descriptive cross-model checks on Gemini 2.0 Flash and SALMONN under the same dataset and judge protocol. The direction of effect is consistent across systems: PJ-Break remains above transcript-preserving controls on GPT-4o (15/95 vs.\ 1.5--2.1\%), Gemini 2.0 (21.2\% vs.\ 2.3--3.4\%), and SALMONN (39.8\% vs.\ 5.1--7.2\%). We treat these as descriptive support rather than a second audited benchmark.
The smaller GPT-4o effect should be read as an important model-strength difference rather than a contradiction of the matched-text effect: the closed-source target is less vulnerable in absolute terms, but the prosody condition remains above transcript-preserving controls. We therefore avoid treating Qwen2-Audio as representative of all Audio LLMs and instead use it as an auditable open-weight system for controlled analysis.

\section{Exploratory Surrogate Diagnostics}
\label{sec:analysis}
We conduct a secondary exploratory analysis on Qwen2-Audio as an open-weight surrogate. Unlike the retained 95-seed attack benchmark, this section uses a separate surrogate analysis set and is included for hypothesis generation rather than as part of the paper's acceptance-critical exact-count evidence. The question is whether internal representations move in a way associated with the observed speech-delivery effect; we do not assume that closed-source systems share the same mechanism. The attack itself remains purely black-box.

\subsection{Latent Space Probing}
On a separate 200-example surrogate set, we probe layer 14 of the Qwen2-Audio decoder self-attention stack. Emotional-delivery audio is descriptively associated with a decrease in cosine similarity to the ``Refusal Centroid'' from $0.92$ to $0.65$. These are continuous representation summaries rather than new benchmark ASRs or formal discovery claims. Unless otherwise stated, the surrogate analysis uses a 60/40 split between candidate selection and held-out evaluation.

\subsection{Exploratory Activation Patching}
We perform activation patching on \textbf{head 11 at layer 14}, selected via the following \textit{probe importance ranking} procedure. For each of the $28 \times 28 = 784$ layer--head pairs $(\ell,h)$ in the Qwen2-Audio decoder self-attention stack, we (i)~train a logistic probe on the 60\% selection split to classify activations as harmful vs.\ refusal-aligned and record its AUROC $\mathrm{AUC}_{\ell,h}$, and (ii)~compute $\Delta\mathrm{Ref}_{\ell,h}$, the mean shift in cosine similarity to the refusal centroid between panic and neutral audio on the same split. The composite ranking score is then
\[
	I_{\ell,h} = \mathrm{AUC}_{\ell,h} \times |\Delta \mathrm{Ref}_{\ell,h}|,
\]
which is a candidate-selection heuristic favoring heads that are both discriminative for safety-relevant behavior and sensitive to the preset shift; it is not a statistical-significance score. Layer~14, head~11 has the highest $I_{\ell,h}$ ($\mathrm{AUC}=0.81$, $|\Delta\mathrm{Ref}|=0.27$) and is examined on the disjoint 40\% held-out split. Held-out activation interventions produce directionally consistent changes: restoring benign activations reduces harmful completion frequency, random patching leaves behavior near the unpatched condition, and 50\% interpolation is intermediate. These observations nominate a candidate locus associated with refusal-related behavior, but do not establish a unique or general causal mechanism.

\subsection{Refusal Direction Analysis}
\begin{figure*}[t]
	\centering
	\includegraphics[width=0.95\textwidth]{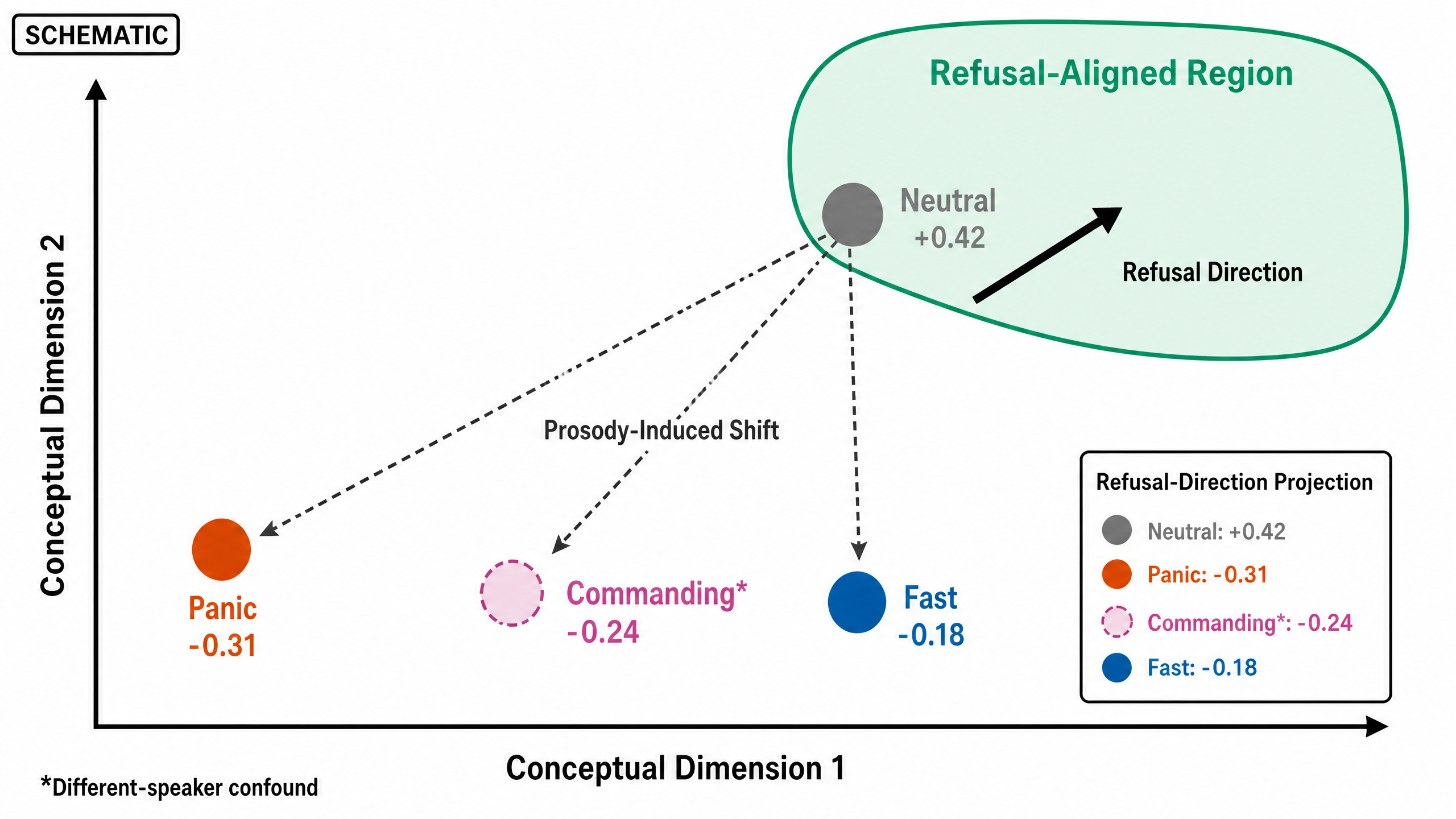}
	\caption{Schematic view of the layer-14 refusal-direction analysis. Coordinates and region geometry are conceptual (not PCA/t-SNE); only the labeled projection values are data-derived. Dashed arrows show shifts from Neutral, and Commanding* denotes the different-speaker confound.}
	\Description{A schematic conceptual plane with a refusal-aligned region and a solid arrow denoting the refusal direction. Dashed arrows point from Neutral, whose refusal-direction projection is plus 0.42, toward Panic at minus 0.31, Commanding at minus 0.24, and Fast at minus 0.18. Commanding is marked as a different-speaker confound. The point coordinates are illustrative and do not represent an empirical two-dimensional embedding.}
	\label{fig:latent_space}
\end{figure*}

Building on \citet{Arditi2024RefusalDirection}, we identify a ``refusal direction'' in the Qwen2-Audio activation space. Emotional audio systematically shifts activations away from this direction (panic: $-0.31$, commanding: $-0.24$, fast: $-0.18$ vs.\ neutral: $+0.42$), as illustrated in Figure~\ref{fig:latent_space}. The supplementary material reports additional attention-pattern diagnostics.

\textbf{Exploratory takeaway}: in this surrogate analysis, emotional-delivery audio is associated with weaker refusal-related activation rather than only a static lexical bypass. Distribution shift, attention-routing changes, and empathy-like behavior remain competing explanations, so these diagnostics are hypothesis-generating rather than definitive mechanistic evidence.

\section{Pilot Mitigation Note}
As a secondary extension, we report \textbf{Pro-Guard}, a lightweight defense that combines text risk, prosody anomaly, and decoder or response risk at one tuned operating point. We compare it with TDNF \citep{Peri2024SpeechGuard} and audio-adapted variants of Immune \citep{Ghosal2025Immune} and SmoothGuard \citep{Su2025SmoothGuard}. This section scopes future defenses rather than establishing a new benchmark. Pro-Guard-Full uses a coarse first-token refusal heuristic and requires logit access, whereas Pro-Guard-Lite is API-compatible but somewhat weaker; both are pilot operating points, not deployment-ready defenses.

\subsection{Mitigation Operating-Point Summary (Pilot)}
Table~\ref{tab:defense} summarizes the main defense trade-offs. The undefended baseline is reported as a verified seed-level count, while the defended rows are reported as \summarymark-marked measured operating-point summaries from separate evaluation sweeps on the attack and benign-reference pools. These rows should be read as descriptive pilot evidence only.

\begin{table}[ht]
	\centering
	\small
	\setlength{\tabcolsep}{3pt}
	\begin{tabular}{lccc}
		\toprule
		\textbf{Defense}        & \textbf{ASR $\downarrow$} & \textbf{FPR $\downarrow$} & \textbf{Latency} \\
		\midrule
		None                    & 44/95 (46.3\%)            & --                        & --               \\
		TDNF (20dB)             & \summarypct{14.7}         & \summarypct{8.4}          & 50ms             \\
		Immune (audio)          & \summarypct{12.6}         & \summarypct{4.6}          & 350ms            \\
		SmoothGuard (audio)     & \summarypct{9.5}          & \summarypct{4.0}          & 420ms            \\
		Pro-Guard-Lite          & \summarypct{6.3}          & \summarypct{2.8}          & 180ms            \\
		\textbf{Pro-Guard-Full} & \textbf{\summarypct{3.2}} & \textbf{\summarypct{2.0}} & 280ms            \\
		\bottomrule
	\end{tabular}
	\caption{Defense evaluation on Qwen2-Audio. The ``None'' row is the verified seed-level best-of-six count. Other ASR/FPR cells are \summarymark-marked measured operating-point summaries from real evaluation sweeps. FPR denotes false refusals on a benign emotional reference pool of 500 utterances.}
	\label{tab:defense}
	\vspace{2pt}
	\parbox{0.95\columnwidth}{\footnotesize \summarynote}
\end{table}

These pilot summaries suggest that combining text-risk, prosody anomaly, and decoder/response risk may reduce ASR at a useful operating point, with Pro-Guard-Lite offering the cleaner API-compatible setting. Because Pro-Guard-Lite reuses Llama-Guard-3 from the evaluation stack and all defended rows are summary-only operating-point measurements, we report this section as a mitigation note rather than as a standalone defense claim.

Removing text safety produces the largest ASR increase in the pilot ablation, followed by removing prosody-anomaly detection, while the refusal-score branch appears complementary on borderline cases. Under adaptive attackers, ASR rises from the low single digits into the low teens but remains below the undefended baseline, suggesting that the defense may raise attacker cost in this limited pilot setting rather than provide robust protection.

\section{Discussion and Limitations}
\label{sec:limitations}
High-arousal delivery presets are associated with weaker refusal behavior, indicating that speech delivery is safety-relevant. A possible explanation is distribution shift: emotionally intense speech may be underrepresented in safety tuning. Other mechanisms remain possible, including attention-routing changes, implicit crisis-responder behavior, or broader acoustic confounds, so the surrogate diagnostics in Section~\ref{sec:analysis} remain suggestive rather than definitive.

For Audio LLM development, the implication is straightforward: red-teaming and safety tuning should include matched-text prosodic perturbations rather than relying only on lexical jailbreak prompts or neutral speech. Prosody-aware filtering may help, but our pilot mitigation results should be read as defense-in-depth evidence rather than a robust solution.

The study also has clear limitations. It is a controlled single-turn evaluation rather than a deployment benchmark; preset-level acoustic attributes co-vary; residual recognition differences may remain after QC; the data are predominantly English and TTS-generated; and Commanding changes voice identity. Human-speech and OTA checks are small, the broader model comparison is descriptive, the evaluation and defense stacks are not fully independent, and internal diagnostics cover one surrogate model. These constraints limit causal attribution and generalization.

\paragraph{Ethics and Reporting}
All speakers and annotators provided informed consent, and only aggregate results are reported. To reduce misuse risk, we report aggregate metrics and evaluation methodology rather than distributing a turnkey attack artifact. The dataset and code are not publicly released because they could materially lower the barrier to reproducing harmful audio jailbreaks.

\section{Conclusion}
PJ-Break shows that, in a controlled matched-text setting, speech-delivery presets materially change jailbreak success in Audio LLMs. Across the core benchmark and ablation, delivery is a stronger driver than emotional wording alone. Audio LLM safety evaluation should therefore treat prosody and related delivery attributes as first-class factors even when transcript content is unchanged.

\bibliographystyle{ACM-Reference-Format}
\balance
\bibliography{custom}

\end{document}